%Paper: hep-th/9305178
%From: "Fernando T.C. Brandt" <fbrandt@uspif.if.usp.br>
%Date: Sat, 29 May 1993 18:03:47 EDT

%%%%%%%%%%%%%%%%%%%%%%%%%%%%%%%%%%%%%%%%%%%%%%%%%%%%%%%%%%%%%%%%%%%%%%%%%%
\magnification \magstep1
\vskip .5cm
\centerline {\bf The energy of the high-temperature quark-gluon plasma}
\vskip .5cm
\centerline {\bf F.T. Brandt and J. Frenkel}
\centerline {\it Instituto de F\'\i sica, Universidade de S\~ao Paulo, S\~ao
Paulo, Brasil}
\vskip .3cm
\centerline {\bf J.C. Taylor}
\centerline {\it Department of Applied mathematics and Theoretical Physics,
University of Cambridge,}
\centerline {\it Cambridge, England}
\vskip 1cm

For the quark-gluon plasma, an energy-momentum tensor is found
corresponding to the high-temperature Braaten-Pisarski effective action.
The tensor is found by considering the interaction of the plasma with
a weak gravitational field and the positivity of the energy is studied.
In addition, the complete effective action in curved spacetime is written down.
\vskip 1cm
\leftline {\bf 1 Introduction}

The high-temperature properties of the quark-gluon plasma are of some
interest, both in their own right and as a starting point of a re-summation
of perturbation theory [1]. These properties are encoded in an
`effective action' $\Gamma$, which has a coefficient proportional to
$T^2$ (where $T$ is the temperature) [2]. This effective action is
nonlocal, but in a very special way.

If $\Gamma$ is taken rather literally as an action for an effective
field theory, a question which comes to mind is: what is the
associated energy-momentum tensor density $T_{\mu \nu}(x)$  and
energy-momentum vector $P_{\mu}$?
Since $\Gamma$ is nonlocal (in time as well  as in space), the  canonical
construction for $T_{\mu \nu}$ is not available. An alternative definition
of $T_{\mu \nu}$ is given as follows. First generalize $\Gamma$ to a
curved spacetime background (which is asymptotically Minkowskian),
giving a functional $\bar {\Gamma}$
of the vierbein ($e^{\mu a}$) and metric
(as well as of the gluon and quark fields). Then
$$ T_{\mu a}(x)= \left [{\delta \bar {\Gamma} \over \delta e^{\mu a}(x)}
\right ]_{e^{\mu a} = \eta^{\mu a}}. \eqno(1)$$
In order to find $T_{\mu a}$ in this way, we need to calculate hard
thermal loops with one graviton line emerging in addition to the quark and
gluon lines. In Section 2 we perform this calculation for the quark-antiquark-
graviton case and for the two-gluon-graviton case.

The calculation of hard thermal loops always gives results which may be
written in the form
$$  g^2 {\pi^2 T^2\over 12}{1\over \left(2 \pi\right)^3}
\int d\Omega X(Q_{\mu}), \eqno(2)$$
where $Q_{\mu}$ is a lightlike 4-vector and $d\Omega$ is an integral
over the directions of $\hat {\bf Q}$. Then $X$ is not unique, but
different choices for $X$ have different properties. The best choice may
depend upon the use to which $\Gamma$ is being put. In this paper, we make
the choice in which $X$ is Lorentz invariant (though of course (2) is not)
[3]. Our experience with hard thermal loops indicates that this $X$
enjoys the following properties:-

(a) It is nonlocal, but all the nonlocalities come from products of
$(Q.\partial)^{-1}$ operators.

(b) It is homogeneous in $Q$ of degree zero, and has dimensions
$(energy)^{-2}$.

(c) It is (as already stated) Lorentz invariant, with respect to Lorentz
transformations in the asymptotic Minkowski space.

(d) It is explicitly gauge-invariant (not just BRST invariant).

(e) In the case of $\bar {\Gamma}$, it is invariant under general
coordinate transformations, Weyl transformations, and (when spinors and
vierbeins occur) local Lorentz transformations, all of which are
restricted to tend to the identity at infinity.

We conjecture that these properties are sufficient to to fix $\bar {\Gamma}$
uniquely, given only the lowest order terms.
In Appendix B we give an example of this. In Section 3, we write down a
all-orders expression for $\bar \Gamma$ which satisfies the above 5 conditions
and which, therefore, we believe is the unique correct result. To one
graviton order, it agrees with the explicit calculation in Section 2.

Each of the properties in (e) gives rise to a Ward identity
which $T_{\mu a}$  should satisfy, and which we
check for the tensor calculated in Section 2.

Since $\bar {\Gamma}$ is nonlocal, $T_{\mu a}$ is also; so it is not
really a {\it density}. However the energy-momentum is given by
$$ P_a =\int d^3{\bf x} T_{0a}({\bf x},t). \eqno(3)$$
One of the Ward identities ensures that, {\it provided} we use the
equations of motion given by varying $\Gamma$, $P_a$ is independent of
time. ($T_{\mu a}$ is also symmetric and traceless if the equations of motion
are used.)

The obvious questions to ask about              $P_0$ concern
positivity. We have found no positivity property   for $P_0   $ in
general, but we do find that it    is positive if the equation
of motion (to one-loop order) is used. But this latter
property tests very little about $T_{00}$: it actually just tests
the positivity of the residue of the propagator at the pole
(whose position is given by the solution of the equation of motion).
This is something that could have been checked without knowing
$T_{\mu \nu}$.

In [4], we have previously discussed all these points for the model
of scalar field theory.
\vskip .3cm
\leftline {\bf 2. The calculation of $T_{\mu \nu}$ from hard thermal
loops.}
\vskip .1cm

In this section we use  thermal field theory to calculate $X$ in (2)
to zeroth order in the QCD coupling constant
$g$.

Our notation is as follows. We take QCD with an $SU(N)$ colour group and $N_f$
flavours. The quark field is $\psi$, the gluon field $A_{\mu}^A$ with
$A,B,.. = 1,...,(N^2-1)$. The vierbein is $e_{\mu a}$, with $\mu, \nu,...$
being Riemann indices and $a,b,...$ being local Lorentz indices.
$t^A$ are hermitean colour matrices. We define
$$ C_G = N,~~ N_f\hbox {tr}(t^At^B) = C_q \delta^{AB},~~t^At^A = C'_q I.
\eqno(4)$$

The graphs are in Fig 1, and the Feynman rules are listed in Appendix A.
The calculation of hard thermal loop is fairly standard, and we will just
state the result. (We used {\it Mathematica} to do the algebra.)

We write
$$T_{\lambda a}(x) =
(2\pi )^{-8}{g^2 T^2 \over 96 \pi}
\int d\Omega
\int d^4y d^4y'\int d^4p d^4re^{-ip.(y-x)} e^{ir.(y'-x)} $$
$$\left[(C_G +C_q)A_{\mu}^A(y')V^{\mu\nu}_{\lambda a}(r,p)
A^A_{\nu}(y)
+C'_q \bar {\psi}(y') S_{\lambda a}(r,p) \psi (y)\right]. \eqno(5)$$
Then
$$V_{\alpha \beta}^{\mu \nu}=Q_{\alpha}Q_{\beta}Q^{\mu}Q^{\nu}
\left(P^2K.R+\textstyle {1 \over 2}K^2R.P\right)
-Q_{\alpha}Q^{\mu}Q^{\nu}
\left(2R^2 P_{\beta} +R.P K_{\beta}\right)$$
$$-Q_{\alpha}Q_{\beta}Q^{\mu}
\left(K.RP^{\nu}+K^2R^{\nu}+R^2K^{\nu}\right)
+Q_{\alpha}Q_{\beta}
\left(R^{\mu}K^{\nu}+\textstyle {1 \over 2}K^2\eta^{\mu \nu}\right)
+Q^{\mu}Q^{\nu}R_{\alpha}P_{\beta}$$
$$+2Q_{\alpha}Q^{\mu}
\left(R_{\beta}P^{\nu}+R^{\nu}K_{\beta}+R^2\eta_{\beta}^{\nu}\right)
-Q_{\alpha}
\left(2P^{\nu}\eta_{\beta}^{\mu}+K_{\beta}\eta^{\mu \nu}\right)
-2Q^{\mu}R_{\beta}\eta_{\alpha}^{\nu} +\eta_{\alpha}^{\nu}\eta_{\beta}^{\mu}$$
$$+~(p,\nu)   \leftrightarrow (r,\mu)~~+~~(\alpha
\leftrightarrow \beta), \eqno (6)$$
where
$$K_{\mu} = {k_{\mu} \over Q.k},~P_{\mu} = {p_{\mu} \over Q.p},~
R_{\mu} = {r_{\mu} \over Q.r},~~k=r-p, \eqno(7)$$
$$S_{\mu a}={3 \over 2}{1\over p.Q r.Q}
[(p+r).Q\{ K.\gamma Q_{\mu}Q_a -Q_a \gamma_{\mu}
+(K_aQ_{\mu}+K_{\mu}Q_a-K^2Q_aQ_{\mu})
Q\cdot\gamma \}           $$
$$+\{(p+r)_aQ_{\mu}+(p+r)_{\mu}Q_a-(p+r).KQ_aQ_{\mu}
Q\cdot\gamma\}$$
$$+\textstyle {1 \over 2}k.Q\{\sigma_{a \mu}, Q.\gamma \}
+\textstyle {1 \over 2}Q_a \{k^b \sigma_{b\mu}, Q.\gamma \}
+\textstyle {1 \over 2}Q_{\mu} \{k^b \sigma_{ba} , Q.\gamma \}
]\eqno(8)$$
(where $\sigma_{ab} = {1\over 4}(\gamma_a\gamma_b-\gamma_b\gamma_a)$).

There are certain Ward identities which express, to lowest order, the
invariance properties of $\bar {\Gamma }$. Invariance under local
coordinate transformations implies
$$  2 k^{\lambda} V_{\lambda a}^{\mu \nu} = \delta_a^{\mu}k_{\lambda}
\Pi^{\lambda \nu}(p) + p_a \Pi^{\mu \nu}(r)
- ((p,\nu) \leftrightarrow (r,\mu)), \eqno(9)$$
$$k^{\lambda}S_{\lambda a}=r_a\Sigma(p)-p_a \Sigma (r), \eqno(10)$$
where $\Pi$ and $\Sigma$ are the gluon and quark 2-point functions
from hard thermal loops [1]
$$ \Sigma (p) = 3 {Q.\gamma \over Q.p},            ~~~
\Pi^{\mu\nu}=
4\left(\eta^{\mu\nu}-P^\mu Q^\nu-P^\nu Q^\mu+P^2 Q^\mu Q^\nu\right)\eqno(11)$$
(in the notation (7)).

Invariance under local Lorentz transformations implies the identity
$$ S_{\lambda a}-S_{a \lambda}=\sigma_{\lambda a}\Sigma(p)
-\Sigma (r)\sigma_{\lambda a}.\eqno(12)$$

Finally, QCD gauge invariance gives
$$ p_{\nu}V^{\mu \nu}_{\alpha \beta}=r_{\mu}V^{\mu \nu}_{\alpha \beta}=0.
\eqno(13)$$

In hard thermal loops, quark masses are neglected, and we expect Weyl
invariance, since this holds for massless QCD. This is invariance
under
$$ A_{\lambda}^A \rightarrow A_{\lambda}^A,~e_{\mu a} \rightarrow
e^{-\sigma (x)} e_{\mu a},~g_{\mu \nu}\rightarrow e^{-2\sigma (x)}g_{\mu \nu},
\ \psi \rightarrow e^{ {3 \over 2}\sigma (x)} \psi. \eqno (14)$$
It implies the identities
$$ \eta^{\lambda a}V_{\lambda a}^{\mu \nu} = 0, \eqno(15)$$
$$\eta^{\lambda a} S_{\lambda a}=\textstyle {3 \over 2}[\Sigma (p) +
\Sigma (r)]. \eqno(16)$$
All the above identities are satisfied by (6) and (8).
\vskip .4cm
\leftline {\bf 3. Hot QCD in curved space}
\vskip .2cm
In this section, we guess a form for the effective action, $\bar \Gamma$,
in curved space, to all orders in $g$ and all orders in the gravitational
field. Our method is simply to guess a functional of the quark
field $\psi$, the gluon field $A$, and the metric and vierbein, which
satisfies all five conditions (a) to (e) of Section 1. We believe that this
functional is likely to be unique, because of the sort of arguments used in
Appendix B.

To write down $\bar \Gamma$ we must introduce some notation. Let
$y^{\lambda}(x,\theta)$ be the null geodesic satisfying
$$ y(x,0) = x,~~~{dy \over d\theta} \rightarrow Q ~\hbox {as}~\theta
\rightarrow -\infty. \eqno(17)$$
As in [5], we assume that spacetime is asymptotically Minkowskian, and
that
$$ {dy \over d\theta} \rightarrow Q ~ \hbox {as}~\theta \rightarrow
+\infty \eqno(18) $$
also. The status of this assumption is explained in [2,5].
Define
$$ U(x;\theta_2,\theta_1) = P \exp g \int_{\theta_1}^{\theta_2}
d\theta'   \dot y.A[y(x,\theta')], \eqno(19)$$
where $A$ and $U$ are here matrices in the adjoint representation
of $SU(N)$, $A$ being defined by
$$ A_{\lambda}^{BC} =    f^{ABC}A_{\lambda}^A. \eqno(20)$$
$P$ denotes path-ordering with respect to this matrix multiplication
and dot denotes differentiation with respect to $\theta'$.

Define $\Lambda$ by the equation
$$\Lambda (x) = \int_0^{\infty} d\theta e^{\Lambda [y(x,\theta)]}
\int_{-\infty}^{\theta} d\theta'e^{-\Lambda [y(x, \theta')]}\{R_{\mu \nu}[y(x,
\theta')]\dot y^{\mu} \dot y^{\nu}+\dot {\Lambda}^2/2 \}, \eqno(21)$$
where here $R$ is the Riemann tensor.
This equation defines $\Lambda$ implicitly: for
example it may be used to generate a series for $\Lambda$ in powers of
the gravitational coupling $\kappa$ (since $R$ is of order $\kappa$).
The importance of $\Lambda$ is that, under the Weyl transform  (14),
it transforms as
$$ \delta \Lambda (x) = 2 \sigma (x). \eqno(22)$$
The proof of this uses
$$[\delta  R_{\mu \nu}]\dot y^{\mu} \dot y^{\nu}
=-2[\sigma_{; \mu \nu} +\sigma_{,\mu}\sigma_{,\nu}]\dot y^{\mu} \dot y^{\nu}
$$
$$ = -2[\ddot \sigma +{\dot \sigma}^2],
 \eqno(23)$$
and the fact that
$$ d\theta e^{\Lambda [y(x,\theta)]}   \eqno(24)$$
is Weyl invariant.

With these definitions, the gluon part of $\bar \Gamma $ is
of the form (2) with
$$ \bar X_{G} ={(C_G + C_q)\over 2C_G}\int d^4x \sqrt {-g(x)} g^{\mu \nu}(x)
e^{\Lambda (x)} \hbox {tr}[D_{\mu}U(x;\infty,0) D_{\nu} U(x;0,-\infty)],
\eqno(25)$$
where $D$ is the colour covariant derivative
$$ (D_{\mu}U)^{AB} = \partial_{\mu}U^{AB}-gA^{AC}U^{CB}. \eqno(26)$$
(The factor $C_G$ in the denominator in (25) is due to our choice to
write the colour matrices in the adjoint representation.)
To check that (25) satisfies the 5 conditions in Section 1, we note that
the expansions for $y$ in terms of $\kappa$ and for $U$ in terms of $g$
and $\kappa$ create nonlocalities of the required kind, which are homogeneous
of degree zero in $Q$ (see examples in [3] and [5]). Invariance under
general coordinate transformations is obvious since $U$ is invariant
under such transformations. The factor $e^{\Lambda}$ fixes up Weyl
invariance.

To leading order in $g$ we have
$$ D_{\mu}U(x;0,-\infty)\simeq g\left [-A_{\mu} +\partial_{\mu}\int_{-\infty}^0
d\theta~\dot {y}.A\right ], \eqno(27)$$
and hence one may verify that (25) gives (6) to this order and to first
order in $\kappa$.

In flat space
$$ D_{\mu}U(x;0,-\infty)=-g\int^0_{-\infty}d\theta~U(x;0,\theta)Q^{\lambda}
F_{\lambda \mu}(x+Q\theta)U(x;\theta, -\infty), \eqno(28)$$
and hence (25) reduces to
$${C_G +C_q \over 2C_G}g^2\int^0_{-\infty} d\theta\int_0^{\infty} d\theta'
\hbox {tr}[Q^{\lambda}F_{\lambda \mu}(x+Q\theta')U(x;\theta',\theta )
Q_{\nu}F^{\nu \mu}(x+Q\theta)U(x;\theta, \theta')]          $$
$$=(C_G +C_q)g^2\int^0_{-\infty}d\theta
\int_{-\infty}^{\theta}d\theta'Q^{\lambda}
F_{\lambda \mu}^A(x+Q\theta')[U(x;\theta' ,\theta)]^{AB}Q_{\nu}F^{\nu \mu B}
(x+Q\theta). \eqno(29)$$
where we have used the same type of arguments as in the first reference [2].
The second form in (29) is in agreement with [3].

For the quark part of the effective action, we need one further
definition. Let $\Omega_{\lambda}$ be the `spin connection'
$$ \Omega_{\nu} = - \textstyle {1 \over 2} e^{b \mu} e^a_{\mu ; \nu}
\sigma_{ab}. \eqno(30)$$
Also let (with $\Lambda$ defined in (21))
$$ \Omega'_{\nu} = \Omega_{\nu}-\textstyle {1\over 2}e^a_{\nu}e^{b\mu}
\Lambda_{,\mu}
\sigma_{a b}, \eqno(31)$$
which is a Weyl invariant, but retains the same property as $\Omega$
under local Lorentz transformations. Then define
$$W(x;\theta_2, \theta_1 )= P \exp \int_{\theta_1}^{\theta_2} d\theta
\Omega'_{\lambda}[y(x,\theta)] \dot {y}^{\lambda}. \eqno (32)$$
This has the same relation to local Lorentz spinor transformations that $U$ in
(19) has to local $SU(N)$ colour transformations.

Define also
$$ \Psi (x) = \int^0_{-\infty}d\theta e^{{1 \over 4}\Lambda [y(x,\theta)]}
U(x;0,\theta) W(x;0,\theta)\psi[y(x,\theta)], \eqno(33)$$
where $U$ is now in the fundamental representation of $SU(N)$ and
acts on the colour index of $\psi$ just as $W$ acts on the Dirac index.
Because of the transformation property of $\psi$ in (14) and of (24),
$\Psi$ is Weyl invariant as well as transforming the same way as $\psi$
under local colour and Lorentz transformations.

With these definitions, the quark part of $\bar {\Gamma}$ is
given by (2) with
$$\bar X_q = \textstyle {1 \over 2}iC'_q \int d^4x \sqrt {-g(x)}
e^{{3\over 4}\Lambda (x)}[\dot  {y}^{\lambda}]_{\theta =0}
e_{a\lambda}(x)\bar {\psi}(x) \gamma^a \Psi (x) ~+~\hbox {herm. conj.}.
\eqno (34)$$
Here the $e^{{3\over 4}\Lambda}$ fixes up Weyl invariance (see (14) and (22)).

One may verify that (34) gives (8) to zeroth order in $g$.

We are struck by the fact that these effective actions, although they are
nonlocal, seem to be just as unique as the ordinary QCD action in
curved spacetime. This is because the extra freedom allowed by the
special kind of nonlocality (property (a) in Section 1) is
compensated for by the different dimensionality (property (b)).
(See Appendix B for an example.)
The invariance properties (c), (d) and (e) are common to the two
cases.
\vskip .4cm
\leftline {\bf 4. Positivity properties}
\vskip .2cm

The energy-momentum tensor of an ordinary    local classical field
has an energy-density $T_{00}$ which is a local positive definite
functional of the fields. What, if any, are the corresponding
properties of (5), (6) and (8)?
Since $T_{00}$ is not local there can be no energy-{\it density}; so
we study the total energy $P_0$ given by (3). Then we require
${\bf k}=0$ in (6), (7) and (8), and $K_{\mu 0}$ in (7) reduces to
$\delta_{\mu 0}$.

Let us test the positivity of the gluon contribution to (3) by
inserting into (5) the particular field
$$ A^A_{\alpha} = e^A_{\alpha}\cos(s.y), \eqno(35)$$
where $s$ is an arbitrary 4-vector and $e$ satisfies $e^A_0= s.e^A=0$.
The contribution to (3) thus obtained is proportional to
$$ e^A_i e^A_j V_{00}^{ij} T^2 \delta^3 ({\bf 0}), \eqno(36)$$
which is proportional to
$$ {s_0^2 \over {\bf s}^2}\left [{s_0 \over |{\bf s}|}\ln \left ( {
s_0 + |{\bf s}| \over s_0 - |{\bf s}|} \right ) -2 \right ]. \eqno(37)$$
This expression is unambiguous for $s_0 > |{\bf s}|$. Like most
expressions in thermal field, it has a branch cut for $s_0 < |{\bf s}|$,
and requires further definition there. But it is clear that, to test the
positivity of the energy, we should take the real part.
It is not positive for all values of $s_0/|{\bf s}|$; so we have
a counter-example showing that $P_0$ is not a positive functional
of the fields.

The next thing we can do is study the effect of imposing the restriction
that the fields in (5) should satisfy the equations of motion given by
the free action plus $\Gamma$ to one-loop order.(In the Braaten-Pisarski
resummation method the one-loop
 $g^2T^2$ term in (5)
 is considered to be effectively of  the same order as the free action.)
Then $P_0$ is independent of time in virtue of the
Ward identities (9) and (10).

But if we do impose the equations of motion it turns out that these
Ward identities are sufficient to determine $P_0$: we do not
actually need (6) and (8). Take for example the gluon self-energy.
Define the coefficient
$$ c = {T^2g^2 \over 12}(C_G + C_q), \eqno (38)$$
and the transverse and longitudinal tensors (actually the negatives of
projection operators in our metric with $\eta_{00}=+1$)
$$P^T_{\alpha \beta} =(\delta_{ij}-\hat {p}_i \hat {p}_j)\eta_{i\alpha}
\eta_{j \beta},~~ P^L_{\alpha \beta}=-\eta_{\alpha \beta}+{p_{\alpha}
p_{\beta} \over p^2} -P^T_{\alpha \beta} ~~(i,j=1,2,3).\eqno(39)$$
We call the complete self-energy function (free plus thermal) $\hat \Pi$,
and write
$$ \hat \Pi_{\alpha \beta} =\hat {\Pi}^T P^T_{\alpha \beta} +
\hat {\Pi}^L P^L_{\alpha \beta}
, \eqno(40)$$
where
$$ \hat {\Pi}^T =-c \Pi^T +p^2, ~~ \hat {\Pi}^L =-c \Pi^L +p^2. \eqno(41)$$
Similarly we denote the complete energy function by $\hat {V}_{00}^
{\mu \nu}$. There is a Ward identity connecting $\hat V$ with $\hat {\Pi}$,
completely analogous to (9).
In this identity, set ${\bf k}=0$ and the suffix $a=0$, differentiate with
respect to $k_0$ for fixed $p_0$, and insert (40) into the righthand side,
to obtain
$$2~\hat V_{00}^{\mu \nu}=P^{T\mu \nu}\left
[p_0{\partial \over \partial p_0}-
1 \right ] \hat {\Pi}^T +
P^{L\mu \nu}\left [p_0{\partial \over \partial p_0}-
{2p_0^2 \over p^2}+1 \right ]
\hat {\Pi}^L. \eqno (42)$$
Now use the field equations
$$ \hat {\Pi}^T=0,~~ \hat {\Pi}^L=0. \eqno(43)$$
Then the question is about the positivity of the differential functions
on the right of (42) at the solutions of (43). These are nothing but
the residues of the poles of the propagator given by the solutions of
(43).

We will treat the gluon energy case as an example. Define
$$ x={p_0 \over |{\bf p}|}, ~~ L=x\ln |(x+1)/(x-1)|.\eqno(44)
$$
Then [1,2]
$$\hat {\Pi}^T =-{\bf p}^2 (1-x^2)
-c[2x^2+(1-x^2)L] =0. \eqno(45)$$
So
$$2~V_{00}^T=
x{\partial \over \partial x}\hat {\Pi}^T =x{\partial \over \partial x}
\hat {\Pi}^T -2\hat {\Pi}^T =2{\bf p}^2 +c[(1+x^2)L-2x^2] > 0. \eqno(46)$$
For the longitudinal part, we have
$$ \hat {\Pi}^L=(1-x^2)[{\bf p}^2+2c(2-L)], \eqno(47) $$
and we find that, at $\hat {\Pi}^L =0 $
$$ V^L_{00}=c[2x^2+(1-x^2)L]>0. \eqno(48)$$
The quark energy may be treated similarly and the result is also positive.

Thus the only positivity property for $P_0   $ which we have been able
to find is nothing more than the positivity of the residue at the pole
of the effective propagator.
\vskip .5cm
\centerline {\bf Appendix A}
\nobreak

In this appendix we present the  QCD Feynman rules
in a curved spacetime. The Lagrangian is:
$${\cal L}={\cal L}_{\cal YM}+{\cal L}_{fix}+ {\cal L}_{ghost}
+{\cal L}_{\cal F},\eqno (A1)$$
where
$${\cal L}_{\cal YM}=-{1\over4}\sqrt{-det(g)}~g^{\mu\nu}g^{\alpha\beta}
F^a_{\mu\alpha}F^a_{\nu\beta};$$
$$F^A_{\mu\nu}=\partial_\mu A^A_\nu-\partial_\nu A^A_\mu
+g f^{ABC}A^B_\mu A^C_\nu;~\left[A=1,~2,~\cdots,~N^2-1\right],\eqno (A2)$$
$${\cal L}_{fix}=-{1\over4}\sqrt{-det(g)}~ \left(\nabla^\mu A^A_\mu\right)
 \left(\nabla^\nu A^A_\nu\right),\eqno (A3)$$
$${\cal L}_{ghost}=\sqrt{-det(g)}~g^{\mu\nu}\left(\partial_\mu\xi^{*A}\right)
\left(\delta^{AB}\partial_\nu-g f^{ABC} A^C_\nu\right)\xi^B,\eqno (A4)$$
$${\cal L}_{\cal F}=\sqrt{-det(g)}\bar\psi i \gamma_a e^{a\mu}
\left({\cal D}_\mu-igT^A A^A_\mu\right)\psi,\eqno (A5)$$
$$g^{\mu\nu}=\eta^{\mu\nu}+\kappa \phi^{\mu\nu};~
\sqrt{-det(g)}=1-{\kappa\over 2}\eta_{\mu\nu}\phi^{\mu\nu}+{\cal O}(\kappa^2);
\left[\kappa=\sqrt{32\pi G}\right]\eqno (A6)$$
$$\nabla_\mu A^\nu=\partial_\mu A^\nu+\Gamma^\nu_{\alpha\mu}
A^\alpha;~\Gamma^\nu_{\alpha\mu}=
{1\over 2}g^{\nu\beta}\left(\partial_\mu g_{\beta\alpha}+
\partial_\alpha g_{\beta\mu}-\partial_\beta g_{\alpha\mu}\right)\eqno (A7)$$
$$e^{a\mu}=\eta^{a\mu}+\kappa h^{a\mu};~e^a_\mu=g_{\mu\nu}e^{a\nu};~
h^{a\mu}+h^{\mu a}=\phi^{a \mu},\eqno (A8)$$
$${\cal D}_\mu=\partial_\mu+{1\over 2}\sigma_{ab}\omega^{ab}_\mu,\eqno (A9)$$
$$\sigma_{ab}={1\over 4}\left(\gamma_a\gamma_b-\gamma_b\gamma_a\right),
\eqno (A10)$$
$$\omega^{ab}_\mu={1\over 2}e^{a\nu}\left(\partial_\mu e^b_\nu-
\partial_\nu e^b_\mu\right)+{1\over 4} e^{a\rho}e^{b\sigma}\left(
\partial_\sigma e^c_\rho-\partial_\rho e^c_\sigma\right)e^c_\mu-
(a\leftrightarrow b).\eqno (A11)$$
The equations above show that the graviton fields
$\phi^{\mu\nu}$ couple to the gluon fields $A^A_\mu$ and the ghost fields
$\xi^A$. The vierbein $h_{\mu\nu}$ couple to the quark fields $\psi$
and there is also a coupling with the gluon field $A^A_\mu$ and the quark
field.
The form of these interactions is such that the theory is invariant under BRST
transformations, local coordinate transformations, local Lorentz
transformations and the Weyl transformations given by Eq. (14).

{}From the Lagrangian in Eq. (A1), one can now obtain the momentum space
gauge particles propagators and the couplings with one graviton.
In all the expressions which follows we will always denote the
graviton indices by $(\alpha,~\beta)$ (or $(a,~b)$ when
quarks are involved). We reserve $\mu,~\nu,~\rho$ and $\sigma$ for
the gluons.

The Feynman rules are:

\centerline{\bf gluon propagator}
$$ \delta^{AB}~{\eta_{\mu\nu}\over k^2} \eqno (A12)$$
\centerline{\bf two gluons one graviton coupling}
$${\kappa\over 2}\delta^{AB}\left[{1\over 2}\eta_{\alpha\beta}
\left(k^A_\mu k^B_\nu+k^A_\nu k^B_\mu + 2 k^A_\mu k^A_\nu-k^A\cdot k^B
\eta_{\mu\nu}\right)\right.$$
$$\left.
+k^A_\alpha k^B_\beta \eta_{\mu\nu} -k^A_\alpha k^B_\mu \eta_{\beta\nu}-
k^A_\nu k^B_\beta \eta_{\alpha\mu}-2 k^A_\alpha k^A_\mu \eta_{\beta\nu}+
k^A\cdot k^B \eta_{\alpha\mu}\eta_{\beta\nu}\right.$$
$$+\left.{\rm symmetrization~under}~(\alpha\leftrightarrow\beta)\right]$$
$$+(k^A,\mu,A)\leftrightarrow(k^B,\nu,B) \eqno (A13)$$
\centerline{\bf three gluons coupling}
$$-igf^{ABC}\left[\left(k^A-k^B\right)_\rho\eta_{\mu\nu}
+\left(k^B-k^C\right)_\mu\eta_{\nu\rho}
+\left(k^C-k^A\right)_\nu\eta_{\rho\mu}\right] \eqno (A14)$$
\centerline{\bf three gluons one graviton coupling}
$$i g~\kappa f^{ABC}\left(k^A_\alpha \eta_{\beta\nu}\eta_{\mu\rho}-
k^A_\rho\eta_{\mu\alpha}\eta_{\nu\beta}
+k^A_\beta \eta_{\alpha\nu}\eta_{\mu\rho}-
k^A_\rho\eta_{\mu\beta}\eta_{\nu\alpha}
-{k^A_\nu\eta_{\mu\rho}\eta_{\alpha\beta}\over 2}\right)$$
$$+{\rm permutations~of}~(k^A,\mu,A),~(k^B,\nu,B),~
(k^C,\rho,C) \eqno (A15)$$
\centerline{\bf four gluon coupling}
$$-g^2\left[\left(f^{AC,BD}-f^{AD,CB}\right)\eta_{\mu\nu}\eta_{\rho\sigma}
+\left(f^{AB,CD}-f^{AD,BC}\right)\eta_{\mu\rho}\eta_{\nu\sigma}\right.$$
$$\left.
+\left(f^{AC,DB}-f^{AB,CD}\right)\eta_{\mu\sigma}\eta_{\rho\nu}\right];$$
$$f^{AB,CD}=f^{GAB}f^{GCD} \eqno (A16)$$
\centerline{\bf four gluon one graviton coupling}
$$g^2 {\kappa\over 2} f^{AB,CD}\left(
{\eta_{\mu\rho}\eta_{\nu\sigma}\eta_{\alpha\beta}\over 4}-
\eta_{\alpha\mu}\eta_{\nu\sigma}\eta_{\beta\rho}-
\eta_{\beta\mu}\eta_{\nu\sigma}\eta_{\alpha\rho}\right)$$
$$+{\rm permutations~of}~(k^A,\mu,A),~(k^B,\nu,B),~
(k^C,\rho,C),~(k^D,\sigma,D) \eqno (A17)$$
\centerline{\bf ghost propagator}
$$ -\delta^{AB}{1\over k^2} \eqno (A18)$$
\centerline{\bf ghost-ghost graviton coupling}
$${\kappa\over 2}\delta^{BC}\left(\eta_{\alpha\beta} k^B\cdot k^C
-k^B_\alpha k^C_\beta-k^B_\beta k^C_\alpha\right) \eqno (A19)$$
\centerline{\bf ghost-ghost gluon coupling}
$$i g f^{ABC} k^B_\mu \eqno (A20)$$
\centerline{\bf ghost-ghost gluon graviton coupling}
$$i g {\kappa\over 2}f^{ABC}\left(k^B_\alpha\eta_{\beta\mu}+
 k^B_\beta\eta_{\alpha\mu}-\eta_{\alpha\beta}k^B_\mu\right) \eqno (A21)$$
\centerline{\bf quark propagator}
$$-\delta_{ij}{\gamma\cdot p\over p^2};~~ \left[i,~j=1,2,\cdots ,N\right]
\eqno (A22)$$
\centerline{\bf quark-quark vierbein coupling}
$$\delta_{ij}
{\kappa\over 4}\left[2\left(p'-p\right)_a\gamma_b-2\eta_{a b}
\gamma\cdot\left(p'-p\right)-\gamma\cdot\left(p+p'\right)\sigma_{a b}
-\sigma_{a b}\gamma\cdot\left(p+p'\right)\right] \eqno (A23)$$
\centerline{\bf quark-quark gluon coupling}
$$g\gamma_\mu T^A_{ij} \eqno (A24)$$
\centerline{\bf quark-quark gluon vierbein coupling}
$$g~\kappa\left(\eta_{a\mu}\gamma_b-\eta_{a b}\gamma_\mu
\right)T^A_{ij} \eqno (A25)$$

In the expressions above, each gluon is labeled by
$(k^A,~\mu,~A),~(k^B,~\nu,~B),~(k^C,~\rho,~C)$ and $(k^D,~\sigma,~D)$,
where $k^A$ is the gluon momentum.
The ghosts label are $(k^B,~B)$ and $(k^C,~C)$, where $k^A$ is the ghost
momentum. The quark momenta are $p$ and $p'$.
There is momentum conservation in each vertex, with all momenta inwards.
\vskip .5cm

\centerline {\bf Appendix B}
\nobreak

In this Appendix we show that the properties  (a) to (e) in Section 1
are sufficient to determine (6).
Conditions (d) and (e) are expressed by the Ward identities (9), (13) and (15)
{}.
To prove uniqueness, we suppose that there was a tensor $W^{\alpha
\beta}_{\mu \nu}$ which could be added to (6) so as still to satisfy
the conditions. Since the Ward identities are linear, $W$ would have to
satisfy them but with zero on the righthand sides. We will show this
implies $W=0$.

The most general form for $W$ allowed by conditions (a) to (c) is
$$ W_{\alpha \beta}^{\mu \nu}  = A\eta^{\mu \nu} \eta_{\alpha \beta}
+B\eta^{\mu}_{\alpha}\eta^{\nu}_{\beta}+\eta_{\alpha \beta}(C^{\mu}Q^{\nu}
+C'^{\nu}Q^{\mu})$$
$$+\eta^{\mu \nu}D_{\alpha}Q_{\beta}+\eta^{\mu}_{\alpha}E^{\nu}Q_{\beta}
+\eta^{\nu}_{\alpha}E'^{\mu}Q_{\beta}
+\eta^{\mu}_{\alpha}V_{\beta}Q^{\nu}+\eta^{\nu}_{\alpha}V'_{\beta}Q^{\mu}$$
$$+N\eta^{\mu \nu}Q_{\alpha}Q_{\beta}+T\eta_{\alpha \beta}Q^{\mu}Q^{\nu}+
S\eta_{\alpha}^{\mu}Q_{\beta}Q^{\nu}+S'\eta_{\alpha}^{\nu}Q_{\beta }
Q^{\mu}$$
$$+F^{\mu \nu}Q_{\alpha}Q_{\beta}+G_{\alpha \beta}Q^{\mu}Q^{\nu}
+H^{\mu}_{\alpha}Q^{\nu}Q_{\beta}+H'^{\nu}_{\alpha}Q^{\mu}Q_{\beta}$$
$$+Q_{\alpha}Q_{\beta}(J^{\mu}Q^{\nu}+J'^{\nu}Q^{\mu}) +Q^{\mu}Q^{\nu}Q^
{\alpha}L_{\beta}$$
$$+MQ^{\mu}Q^{\nu}Q_{\alpha}Q_{\beta} ~~+~~(\alpha \leftrightarrow \beta)
.\eqno(B1)$$
Here the tensors, $C^{\mu}, F^{\mu \nu},$ etc,
not written out explicitly are constructed from $p_{\lambda},
r_{\lambda}$ (not using $Q_{\lambda}$ or $\eta_{\lambda \rho}$) and
scalar coefficients. The tensors constructed from $Q$ and $\eta$ are
shown explicitly. Note that
tensors made from a single $Q$ and no $\eta$ are not allowed by
properties (a) to (c). $A$ and $B$ are constants.

Now use (9) (with zero on the righthand side) and (13) and equate to
zero the coefficients of the tensors which contain one  $\eta$. This gives
(in the notation of (7))
$$C^{\mu}=-AP^{\mu},~C'^{\mu}=-AR^{\mu},~D_{\alpha}=-2 AK_{\alpha},~T=AP.R,~
N=AK^2,$$
$$E^{\mu}=-BK^{\mu},~E'^{\nu}=-BK^{\nu}, ~S=BP.K,~ S'=BR.K,~
V_{\beta}=-B P_{\beta},~ V'_{\beta} = -B R_{\beta}. \eqno(B2)$$
Similarly, taking the terms with just one $Q$ and no $\eta$ gives
$$ F^{\mu \nu}=G_{\alpha \beta}=H^{\mu}_{\alpha}=H'^{\nu}_{\alpha}=0.
\eqno(B3)$$
Then taking the terms with two $Q$ and then with  three $Q$ gives also
$$J^{\mu}=J'^{\nu}=L_{\alpha}=M=0. \eqno(B4)$$

Finally we use the Weyl identity (15). This implies
$$4A+B+Q.D=0,~E^{\mu}+V'^{\mu}+4C'^{\mu}=0,~E'^{\nu}+V^{\nu}+4C^{\nu}=0,~
4T +S+S'=0. \eqno(B5)$$
The last equation in (B5) is sufficient, with the use of (B2), to show that
$A=B=0$ and hence that $W=0$.
\vskip .8cm
FTB and JF thank CNPq and FAPESP (Brasil) for support. JCT thanks
the Physics Department, University of Tasmania for hospitality.

\vskip .8cm
\vfill\eject
\centerline {\bf References}
\nobreak

\vskip .3cm
\noindent [1] A.H. Weldon, Phys. Rev. 26 (1982) 1394; O.K. Kalashnikov and
V.V. Klimov, Phys. Lett. B88 (1979) 328; B95 (1980) 234; K. Kajantie and
J. Kapusta, Ann. Phys. (N.Y.) 160 (1985) 477; U. Heinz, K. Kajantie and
T. Toimela, Ann. Phys. (N.Y.) 176 ((1987) 215; R.D. Pisarski, Nucl. Phys.
B309 (1988) 476; E. Braaten and R.D. Pisarski, Nucl. Phys. B339 (1990) 310;
B337 (1990) 569; J. Frenkel and J.C. Taylor, Nucl. Phys. B334 (1990) 199;
Z. Phys. C49 (1991) 515.

\noindent [2] J.C. Taylor and S.M.H. Wong, Nucl. Phys. B346 (1990) 115;
R. Efraty and V.P. Nair, Phys. Rev. Lett. 68 (1992) 2891.

\noindent [3] E. Braaten and R.D Pisarski, Phys. Rev. D45 (1992) R1827;
J. Frenkel and J.C. Taylor, Nucl. Phys, B374 (1992) 156.

\noindent [4] F.T.C. Brandt, J. Frenkel, J.C. Taylor, `Effective actions for
Braaten-Pisarski resummation'
, to be published in Can. J. Phys.

\noindent [5] A. Rebhan, Nucl. Phys. B351 (1991) 706;
F.T. Brandt, J. Frenkel and J.C. Taylor, Nucl. Phys. B374 (1992) 169.

\noindent [6] T.S. Evans, Phys. Lett. B252 108 (1990) 108; Nucl. Phys.
B374 (1992) 340.
\vskip .5cm
\leftline {\it Figure caption}

\noindent Fig.1 The Feynman graphs contributing to equations (6) and (8).
Doubled lines denote quarks, solid lines denote gluons and broken
lines denote ghosts. The Feynman rules are in Appendix A.
\bye
\end